\documentstyle[prl,aps,twocolumn]{revtex}
\begin{document}
\draft

\title{Thermodynamics and Excitation Spectrum of a Quasi-One-Dimensional
Superconductor in a High Magnetic Field }
\author{N. Dupuis}
\address{Laboratoire de Physique des Solides,
Universit\' e Paris-Sud, 91405 Orsay, France }
\date{April 1994}
\maketitle
\begin{abstract}
At high magnetic field, the semiclassical approximation which underlies
the Ginzburg-Landau theory of the mixed state of type II
superconductors breaks down. In a quasi-1D superconductor with an {\it open
Fermi surface}, a high magnetic field stabilizes a cascade of
superconducting phases which ends in a strong reentrance of the
superconducting phase. From a microscopic mean-field model, we determine the
thermodynamics and the excitation spectrum of these quantum superconducting
phases.
\end{abstract}
\narrowtext

It has recently been proposed that quasi-one-dimensional superconductors
should exhibit an unusual phase diagram in a high magnetic field
\cite{Lebed86,Dupuis93}. In a quasi-1D conductor (weakly coupled chains
system) with an {\it open Fermi surface}, the magnetic field does not
quantize the semiclassical orbits which are {\it open} but it induces a
dimensional crossover in the sense that it tends to confine the wave
functions along the chains. This quantum effect of the field strongly
modifies the phase diagram predicted by the
Ginzburg-Landau-Abrikosov-Gor'kov (GLAG) theory which is based on the
semiclassical
phase integral (or eikonal) approximation \cite{Parks69}. The usual
Ginzburg-Landau (GL) regime is followed, when the field is increased, by a
cascade of superconducting phases separated by first order transitions which
ends in a strong reentrance of the superconducting phase where the chains
interact by Josephson coupling. The superconducting state evolves from an
Abrikosov vortex lattice in the GL regime towards a Josephson vortex lattice
in the reentrant phase. Between these two limits, the amplitude of the order
parameter
and the current distribution show a symmetry of a laminar type while the
vortices
still describe a triangular lattice. The cascade of phase transitions
originates in commensurability effects between the periodicity of the order
parameter and the crystalline lattice spacing. This
high-field-superconductivity can survive even in the presence of Pauli pair
breaking because the quasi-1D Fermi surface allows one to construct a
Larkin-Ovchinnikov-Fulde-Ferrell state which can exist far above the
Pauli limited field. A very
important aspect is that the temperature and magnetic field scales are
determined by the coupling between chains \cite{Dupuis93}. This means that
the temperature
and magnetic field ranges where high-field-superconductivity is expected can
be experimentally accessible if  appropriate  (i.e. sufficiently
anisotropic) materials are chosen.

In this letter, we derive the thermodynamics and the excitation spectrum of
these quantum superconducting phases from a microscopic model in the
mean-field approximation. We consider a
strongly anisotropic superconductor described by the dispersion law ($\hbar
=k_B=1$ throughout the paper) $E({\bf k})=v(\vert k_x \vert -k_F)+t_z \cos
(k_zc)$ where the Fermi energy is chosen as the origin of the energies.
$v$ is the Fermi velocity for the motion along the chains and $t_z$ is the
coupling between chains separated by the distance $c$. For a linearized
dispersion law, the $y$ direction parallel to the magnetic field does not
play any role (as long as the Cooper pairs are formed with states of
opposite momenta in this direction) so that we restrict ourselves to a 2D
model. We assume that the zero field critical tempeature $T_{c0}$ is smaller
than $t_z$ so that the superconductor is really 3D: in the GLAG
description, there is no Josephson coupling between chains even at $T=0$.
We consider singlet pairing but we neglect the Zeeman term (i.e. we put the
$g$ factor equal to zero). The
Pauli pair breaking effect can easily be incorporated in the present
description.

In the gauge ${\bf A}(Hz,0,0)$, the one-particle Hamiltonian obtained from
the Peierls substitution ${\cal H}_0=E({\bf k} \rightarrow -i{\bf
\nabla }-e{\bf A})$ is given by
\begin{equation}
{\cal H}_0^\alpha =v(-i\alpha \partial _x -k_F)+\alpha \hat m \omega _c
+t_z\cos (-ic\partial _z)   \,,
\label{Hami}
\end{equation}
where $\alpha =+ \,(-)$ labels the right (left) sheet of the Fermi surface
and $\hat m$ is the (discrete) position operator in the $z$ direction. We
have introduced the energy $\omega _c=Gv$ where $G=-eHc$ is a magnetic wave
vector and $H$ the external magnetic field perpendicular to the system.
${\cal H}_0$
can easily be diagonalized by noting that the momentum along the
chains is a good quantum number and by taking the Fourier transform with
respect to $m$. The eigenstates and the corresponding eigenenergies are
\begin{eqnarray}
\phi _{k_x,l}^\alpha ({\bf r})&=&{1 \over \sqrt{cL_x}} e^{ik_x x}
J_{l-m}(\alpha \tilde t)\,, \\
\epsilon _{k_x,l,\sigma }^\alpha &=&v(\alpha k_x-k_F)+\alpha l\omega _c  \,,
\label{spectre}
\end{eqnarray}
where ${\bf r}=(x,m)$, $\tilde t=t_z/\omega _c$, $\alpha ={\rm sgn} (k_x)$
and $J_l$ is the $l$th
order Bessel function. $L_x$ is the length of the system in the $x$
direction. The spectrum consists in a discrete set of 1D
spectra. The state $\phi _{k_x,l}^\alpha $ is localized around the $l$th
chain with a spatial extension in the $z$ direction of the order of
$\tilde t c$ which corresponds to the amplitude of the semiclassical orbits
\cite{Dupuis93}. Note that the states $\phi _{k_x,l}^\alpha $ can be
obtained from the localized states introduced by Yakovenko
\cite{Yakovenko87} by a gauge transformation.
The superconducting instability can be qualitatively understood
from the spectrum (\ref{spectre}). In zero-field, time-reversal
symmetry ensures that $E_{\uparrow }({\bf k})=E_{\downarrow }(-{\bf k})$ so
that the pairing at zero total momentum present the usual (Cooper)
logarithmic singularity which results in an instability of the metallic
state at a finite temperature $T_{c0}$. A finite magnetic field breaks down
time-reversal symmetry. Nonetheless, we still have $\epsilon
_{k_x,l_1,\uparrow }^\alpha =\epsilon _{q_x-k_x,l_2,\downarrow }^{-\alpha }$
for $q_x=-(l_1+l_2)G$. Thus, whatever the value of the field, some pairing
channels will present the Cooper singularity $\sim \ln (2\gamma \Omega /\pi
T)$ ($\gamma \sim 1.781$ and $\Omega $ is the cutoff energy of the
attractive interaction) if the total momentum along the chain $q_x$ is a
multiple of $G$. This results
in logarithmic divergences at low temperature in the linearized gap equation
which destabilize
the metallic state at a temperature $0<T_c<T_{c0}$ \cite{Lebed86,Dupuis93}.
Besides the most singular channels which present the Cooper singularity,
there exist less singular channels with singularities $\sim \ln \vert \Omega
/n\omega _c \vert $ ($n\ne 0$) for $T\ll \omega _c$. In this high field
limit ($\omega _c \gg T$),
a natural approximation consists in retaining only the most singular
channels. Such an approximation
has been used previously in the mean-field theory
of the field-induced-spin-density-wave phases observed in some
organic conductors (Single Gap Approximation)
\cite{FISDW} and in the mean-field theory of isotropic superconductors in a
high
magnetic field where it is known as the Quantum Limit Approximation (QLA)
\cite{Tesanovic92}. In the following, we shall adopt this latter
designation.

In order to obtain the critical temperature when $\omega _c \gg T$, we
consider the two-particle vertex function
in the representation of the eigenstates of ${\cal H}_0$. In the
chosen gauge, the total momentum $q_x$ along the chains is a constant of
motion. Since in the QLA we consider only the most singular
pairing channels, the center of gravity of the Cooper pair in the
perpendicular direction is related to the total momentum $q_x$ by
$L_{12}=(l_1+l_2)/2=-q_x/(2G)$ as explained above, and therefore becomes
also a constant of motion.
Thus the two-particle vertex function can be written as $\Gamma
_{q_x(L),L}^{\alpha
\alpha '}(l_{12},l_{12}')$ with $q_x(L)=-2LG$, where $l_{12}=l_1-l_2$ and
$l_{12}'=l_1'-l_2'$
describe the relative motion of the pair in the $z$ direction. In the
ladder approximation, the integral
equation for $\Gamma $ then reduces to ($l\equiv l_{12}$, $l'\equiv
l_{12}'$)
\begin{eqnarray}
\Gamma _{q_x(L),L}^{\alpha \alpha '}(l,l') &=& -\lambda
V_{l,l'}^{\alpha \alpha '} \nonumber \\
& & +{\lambda \over 2} \chi (0) \sum _{\alpha '',l''}
V_{l,l''}^{\alpha \alpha ''}
\Gamma _{q_x(L),L}^{\alpha ''\alpha '}(l'',l')  \,,
\end{eqnarray}
where $V_{l,l'}^{\alpha \alpha '}=\alpha ^l {\alpha '}^{l'}V_{l,l'}$ and
\begin{equation}
V_{l,l'} =
\int _0^{2\pi } {{dx} \over {2\pi }}
J_l(2\tilde t\cos x)
J_{l'}(2\tilde t\cos x)   \,.
\end{equation}
$-\lambda V_{l,l'}^{\alpha \alpha '}$
is the matrix element of the local electron-electron interaction
$-\lambda \delta ({\bf r}_1-{\bf r}_2)$ ($\lambda >0$) in the representation
of the states $\phi _{k_x,l}^\alpha $. Note that $V_{l,l'}^{\alpha \alpha
'}$ is independent of the center of gravity of the Cooper pair.
$\chi (0)=N(0) \ln (2\gamma \Omega /\pi T)$ is the pair susceptibility at
zero total momentum in zero field and $N(0)$ is the density of states per
spin at the Fermi level.
The preceding integral equation is solved by introducing the
orthogonal transformation $U_{l,l'}$ which
diagonalizes the matrix $V_{l,l'}$. One obtains:
\begin{equation}
\Gamma _{q_x(L),L}^{\alpha \alpha '}(l,l')=-\lambda
\alpha ^l {\alpha '}^{l'} \sum _{l''}
{{U_{l,l''} \bar V_{l'',l''} (U^{-1})_{l'',l'} }
\over {1-\lambda \chi (0) \bar V_{l'',l''} }}  \,,
\label{vertex}
\end{equation}
where $\bar V$ is the diagonal matrix $U^{-1}VU$. The metallic state becomes
instable when a pole appears in the two-particle vertex function which leads
to the critical temperature
\begin{equation}
T_c={{2\gamma \Omega } \over {\pi }} e^{{-1} \over {\lambda N(0) \bar
V_{l_0,l_0}}}  \,,
\end{equation}
where $\bar V_{l_0,l_0}$ is the highest eigenvalue of the matrix $V$.
The critical temperatures are shown in Fig.\ref{Fig1} for the two highest
eigenvalues of $V$. This clearly indicates that there are two lines of
instability competing with each other and leading to a cascade of first
order transitions in agreement with the exact mean-field calculation of $T_c$
\cite{Dupuis93}. Except for the last phase, $T_c$ calculated in the QLA is
several orders of magnitude below the exact critical temperature:
it has been pointed out previously that the QLA strongly
underestimates the critical temperature \cite{Montambaux88}.
The existence of two lines of instability results
from the fact that $V_{l,l'}=0$ if $l$ and $l'$ do not have the same parity.
Diagonalizing the matrix $V_{l,l'}$ is then equivalent to separately
diagonalizing the matrices $V_{2l,2l'}$ and $V_{2l+1,2l'+1}$. In the
following,
we label these two lines by $l_0=0,1$ so that $\bar V_{l_0,l_0}= \max _l
\bar V_{2l+l_0,2l+l_0}$. Since $2L=l_1+l_2$ and $l_1-l_2$ have the same
parity,
$L$ is integer (half-integer) for $l_0=0$ ($l_0=1$) and can be written as
$L=-l_0/2+p$ with $p$ integer. Correspondingly, we have $q_x(L)=(l_0-2p)G$.
It is clear that the instability line $l_0$ corresponds to the instability
line $Q=l_0G$ which was previously obtained in another approach
\cite{Dupuis93}.
{}From Eq.(\ref{vertex}), one can see that the superconducting condensation
in the channel $q_x(L),L,l_0$ corresponds to the following spatial
dependence for the order parameter
\begin{equation}
\Delta _{q_x(L),L,l_0}({\bf r})\sim \sum _{\alpha ,l} \alpha ^l U_{l,l_0}
\phi _{k_x,L+{l \over 2}}^\alpha ({\bf r})
\phi _{q_x(L)-k_x,L-{l \over 2}}^{\overline \alpha }({\bf r})  \,,
\label{po1}
\end{equation}
where we note $\overline \alpha =-\alpha $.
Noting that the matrices $V$ and $U$ have a range of the order of $\tilde t$
(i.e. $V_{l,l'}$, $U_{l,l'}$ are important for $\vert l \vert, \, \vert l'
\vert < \tilde t$), one can see that $\Delta _{q_x(L),L,l_0}$ has the form
of a strip extended in
the direction of the chains and localized in the perpendicular direction on
a length of the order of $c\tilde t$. This is not surprising since
$\Delta _{q_x(L),L,l_0}$ results from pairing between the localized states
$\phi _{k_x,l}^\alpha $.

Following the original approach
proposed by Abrikosov \cite{Parks69}, we construct the order
parameter for $T<T_c$
as a linear combination of the solutions (\ref{po1}):
$\Delta ({\bf r})=\sum _L \gamma (L) \Delta _{q_x(L),L,l_0}({\bf r})$
where $2L$ must have the parity of $l_0$. Since $\Delta _{q_x(L),L,l_0}$ is
localized in the $z$ direction with an extension of the order of $c\tilde
t$, a natural choice for the coefficients $\gamma (L)$
is to take $\gamma (L)\ne 0$ if $L=-l_0/2 +pN'$ ($p$ integer) where the
unknown
integer $N'$ is expected to be of the order of $\tilde t$. In order to
correctly describe the triangular Josephson vortex lattice in the last phase
($\tilde
t \ll 1$) \cite{Dupuis93}, we choose $\gamma (L)\equiv \gamma (p)=1\,(i)$
for $p$ even (odd) which leads to (noting $N=2N'$)
\begin{eqnarray}
\Delta _{l_0,N} ({\bf r}) &=& \Delta \sum _{l,p} U_{2l+l_0,l_0} \gamma _p
e^{i(l_0-pN)Gx} J_{p {N \over 2}+l-m} (\tilde t) \nonumber \\
& & \times J_{p {N \over 2}-l-l_0-m} (-\tilde t)  \,,
\label{po2}
\end{eqnarray}
where the amplitude $\Delta $ is chosen real. Eq.(\ref{po2}) defines a
variational order parameter where the two unknown parameters $\Delta $ and
$N$ have to be determined by minimizing the free energy. It can be seen
that $\vert \Delta ({\bf r})\vert $ has periodicity $a_x=2\pi /NG$ and
$a_z=Nc$ so that the unit cell contains two flux quanta: $Ha_xa_z=2\phi _0$
(when a triangular lattice is described with a square unit cell, the unit
cell contains two flux quanta). In Ref.\cite{Dupuis93}, the order parameter
was constructed by imposing that it describe both the triangular Abrikosov
vortex lattice in weak field ($\omega _c\ll T$) and the triangular Josephson
vortex lattice in very strong field ($\omega _c \gg t_z$). Both approaches
lead to the same order parameter when only the Cooper singularities are
retained.

In order to derive the thermodynamics and the excitation spectrum in the
superconducting phases, it is necessary to determine the normal and
anomalous Green's functions from the Gor'kov equations
\begin{eqnarray}
& & (i\omega -{\cal H}_0^\alpha ) G_\sigma ^\alpha ({\bf r},{\bf
r}',\omega ) -\Delta _\sigma ({\bf r})
F_{\overline \sigma }^{\overline \alpha \dagger }({\bf r},{\bf r}',\omega )
= \delta ({\bf r}-{\bf r}') \,, \nonumber \\
& & (-i\omega -{\cal H}_0^{\overline \alpha \dagger })
F_{\overline \sigma }^{\overline \alpha \dagger }({\bf r},{\bf r}',\omega )
+\Delta _\sigma ^*({\bf r})
G_\sigma ^\alpha ({\bf r},{\bf r}',\omega ) = 0 \,,
\label{Geq}
\end{eqnarray}
with the self-consistency equation
$\Delta _\sigma ^*({\bf r}) = \lambda T \sum _{\alpha ,\omega }
F_\sigma ^{\alpha \dagger }({\bf r},{\bf r},\omega )$. Here $\Delta _\sigma
({\bf r})=\lambda \sum _\alpha \langle \psi _\sigma ^\alpha ({\bf r})  \psi
_{\overline \sigma }^{\overline \alpha }({\bf r}) \rangle $ where the $\psi
_\sigma ^\alpha ({\bf r})$'s are fermionic operators for particles moving on
the sheet $\alpha $ of the Fermi surface. $\Delta _\uparrow ({\bf r})=
-\Delta _\downarrow ({\bf r})$ is the variational order parameter defined
by (\ref{po2}). $G_\sigma ^\alpha $ and $F_\sigma
^{\alpha \dagger }$ are the Fourier transforms with respect to the imaginary
time $\tau $ of the correlation functions $-\langle T_\tau \psi _\sigma
^\alpha ({\bf r},\tau ) \psi _\sigma ^{\alpha \dagger }({\bf r},0) \rangle $
and $-\langle T_\tau \psi _\sigma ^{\alpha \dagger }({\bf r},\tau )
\psi _{\overline \sigma }^{\overline \alpha \dagger }({\bf r},0) \rangle $.
In (\ref{Geq}), it is assumed that the magnetization ${\bf M}=({\bf B}-{\bf
H})/4\pi $ is equal to zero. This approximation is justified in the
last phase ($\tilde t\ll 1$) where ${\bf M}$ is of the order of $\tilde t^2$
and its
contribution to the Gibbs free energy $G(T,H)$ of the order of $\tilde t^4$.
In the other phases, we expect that the approximation ${\bf B}={\bf H}$
will give reliable results at least not too far from the reentrant phase.

In order to have a simple description of the superconducting state, we
introduce the magnetic Bloch states
\begin{equation}
\phi _{{\bf q},l}^\alpha = \sqrt{{N \over {N_z}}}
\sum _p e^{-ipq_za_z} \phi _{q_x+pNG,l-pN}^\alpha \,,
\end{equation}
where $N_z$ is the number of chains. The eigeneneries are $\epsilon
_{{\bf q},l,\sigma }^\alpha =\epsilon _{q_x,l,\sigma }^\alpha $. ${\bf q}$
is restricted to the first magnetic Brillouin zone:
$\vert q_x \vert -k_F \in \rbrack -\pi /a_x,\pi /a_x \rbrack $ and $q_z \in
\rbrack -\pi /a_z ,\pi /a_z \rbrack $. There are $N$ branches
($-N/2< l\le N/2$)
at the Fermi level. The order parameter (\ref{po2}) is entirely
described by the pairing between the states $\phi _{{\bf q},l}^\alpha $ and
$\phi _{l_0{\bf G}-{\bf q},-l_0-l}^{\overline \alpha }$ where ${\bf
G}=(G,0)$. Consequently, the Gor'kov equations (\ref{Geq}) become diagonal
in the representation of the magnetic Bloch states and their solutions are
\begin{eqnarray}
G_\sigma ^\alpha ({\bf q},l,\omega )&=&
{{-i\omega -\epsilon  _{{\bf q},l,\sigma }^\alpha } \over
{\omega ^2+{\epsilon  _{{\bf q},l,\sigma }^\alpha }^2 + \vert \Delta _\sigma
^\alpha ({\bf q},l) \vert ^2 }} \,,  \nonumber \\
F_\sigma ^{\alpha \dagger }({\bf q},l,\omega )&=&
{{\Delta _\sigma ^{\alpha }({\bf q},l)^* }  \over
{\omega ^2+{\epsilon  _{{\bf q},l,\sigma }^\alpha }^2 + \vert \Delta _\sigma
^\alpha ({\bf q},l) \vert ^2 }}  \,,
\label{green}
\end{eqnarray}
where the pairing amplitude $\Delta _\sigma ^\alpha ({\bf q},l)$ is defined
by
\begin{eqnarray}
\Delta _\sigma ^\alpha ({\bf q},l) &=&
\int d^2 {\bf r}\, \phi _{{\bf q},l}^\alpha ({\bf r})^*
\phi _{l_0{\bf G}-{\bf q},-l_0-l}^{\overline \alpha }({\bf r})^*
\Delta _\sigma ({\bf r})  \nonumber \\
&=& \Delta _\sigma \alpha ^{l_0} \bar V_{l_0,l_0}
\sum _p \gamma _p e^{-ipq_za_z} U_{2l+l_0+pN,l_0} \,.
\label{amplitude}
\end{eqnarray}
Here $\Delta _\uparrow =-\Delta _\downarrow =\Delta $.
The functions $G_\sigma ^\alpha $ and $F_\sigma ^{\alpha \dagger }$
appearing
in (\ref{green}) are the Fourier transforms of the correlation functions
$-\langle T_\tau b_{{\bf q},l,\sigma }^\alpha (\tau )b_{{\bf
q},l,\sigma }^{\alpha
\dagger }(0) \rangle $ and $-\langle T_\tau b_{{\bf q},l,\sigma }^{\alpha
\dagger }(\tau ) b_{l_0{\bf G}-{\bf q},-l_0-l,{\overline \sigma
}}^{\overline \alpha \dagger
}(0) \rangle $ where $b_{{\bf q},l,\sigma }^\alpha $  ($b_{{\bf q},l,\sigma
}^{\alpha \dagger
} $) are annihilation (creation) operators of a particle with spin
$\sigma $ in the state $\phi _{{\bf q},l}^\alpha $.

{}From the knowledge of the Green's functions, we can calculate the free
energy of the system. Close to $T_c$, the difference between the free
energies of the superconducting and normal states can be obtained in a GL
expansion $F_N \lbrack \Delta \rbrack =A \Delta ^2 +B \Delta ^4/2$
where
\begin{eqnarray}
A &=& {{\lambda ^{-1}} \over S} \int d^2{\bf r}\,
\Bigg \vert
{ {\Delta ({\bf r})} \over \Delta } \Biggr \vert ^2
- {T \over S} \sum _{{\bf q},l,\omega  }
{{\vert \Delta _\sigma ^\alpha ({\bf q},l) / \Delta  \vert ^2  }
\over {\omega ^2+{\epsilon  _{{\bf q},l,\sigma }^\alpha }^2 }}
\,, \nonumber \\
B &=& {T \over S} \sum _{{\bf q},l,\omega  }
{{\vert \Delta _\sigma ^\alpha ({\bf q},l) / \Delta  \vert ^4  }
\over {(\omega ^2+{\epsilon  _{{\bf q},l,\sigma }^\alpha }^2)^2 }}  \,,
\end{eqnarray}
where $S=L_xN_zc$ is the area of the system. The two preceding equations can
be further simplified by using (\ref{amplitude}).
Minimizing the free energy $F_N \lbrack \Delta \rbrack $ with respect to
$\Delta $, we obtain $F_N=-A^2/2B$.
In the reentrant phase where the approximation ${\bf B}={\bf H}$ is
justified, we find that the minimum of $F_N$ is obtained for
$N=2$ \cite{ND}. When the field is decreased from its value in the reentrant
phase, the
system undergoes a first order phase transition and the minimum of $F_N$ is
then obtained for $N=4$. This result is in agreement with
Ref.\cite{Dupuis93} where it is argued that the first order phase
transitions
are due to commensurability effects between the crystalline lattice spacing
and the periodicity of the order parameter.
Unlike what was expected \cite{Dupuis93}, the best value of $N$ switches to
$6$ before
reaching the next first order transition. This indicates the
importance of the screening of the external field  in calculating the free
energy in the phases $N\ge 4$.

The specific heat jump at the transition is obtained from $\Delta C=-T
\partial ^2 F/\partial T^2$. The ratio $\Delta C/C_N$ where $C_N$ is the
specific heat of the normal state is found to be always smaller than the
(BCS) zero-field value and discontinuous at the first order phase
transitions (Fig.\ref{Fig2}). The discontinuity can be related to the slope
$\Delta T/\Delta H$ of the first order transition line \cite{Poilblanc88}:
the slope
is positive for the last transition and negative for the other transitions.
The magnetization is obtained from
$M=-\partial F_N/\partial H$. Since $T_c$ is determined by $F_N=0$, $M$ has
the sign of $dT_c/dH$ \cite{Montambaux89}. Each phase will therefore first
be paramagnetic and then diamagnetic for increasing field, except the
reentrant phase which is always paramagnetic \cite{Melo}.

{}From the Green's functions (\ref{green}), we deduce the quasi-particle
excitation spectrum $E_{{\bf q},l,\sigma }^\alpha =\pm ({\epsilon _{{\bf
q},l,\sigma }^\alpha }^2
+\vert \Delta _\sigma ^\alpha ({\bf q},l)\vert ^2 )^{1/2}$.
A gap $2\Delta _\sigma ^\alpha ({\bf q},l)$ opens at the Fermi level in each
branch $l$.
The spectrum is shown in Fig.\ref{Fig3} for the last three phases $N=2$,
$N=4$ and $N=6$. In the very high field limit ($\tilde t\ll 1$), the
spectrum is almost flat,
the dispersion of the quasi-particle band being of the order of $\tilde
t^2$. When the field is decreased within a given phase, the dispersion
increases. The minimum excitation energy decreases when $N$ increases.
Thus Fig.\ref{Fig3} clearly shows how the system evolves from a quasi-1D
(quasi-2D if the magnetic field direction is taken into account) behavior
in very high magnetic field ($\tilde t\ll 1$) towards the GL regime ($\omega
_c \ll T$) where the spectrum is known to be gapless \cite{Parks69}.

In conclusion, we have solved the BCS theory for a quasi-1D superconductor
in a high magnetic field. The theory can easily be extended to
include the pairing channels which are not considered in the QLA: the
results presented in this letter are not qualitatively modified.

The author would like to thank G. Montambaux for a critical reading of the
manuscript and a constant interest in this work.

\begin{figure}
\caption{Solid lines: critical temperature vs magnetic field for $l_0=0$ and
$l_0=1$ in the QLA. Dashed line: exact mean-field critical temperature. }
\label{Fig1}
\end{figure}

\begin{figure}
\caption{Ratio $r=(\Delta C/C_N)/(\Delta C/C_N)_{BCS}$ vs magnetic field. }
\label{Fig2}
\end{figure}

\begin{figure}
\caption{Excitation spectrum in the phases $N=2$, $N=4$ and $N=6$. The
units are chosen so that $\max \vert \Delta _\sigma ^\alpha (q_z,l) \vert
=1$. In the phase $N$, there are $N/2$ distinct branches. }
\label{Fig3}
\end{figure}


\begin{references}
\bibitem{ND} We have verified that the square lattice (with periodicity
$N=1$) has a higher energy than the triangular lattice in the reentrant
phase.
\bibitem{Melo} This paramagnetic/diamagnetic crossover as the field is
increased within a given phase has also been proposed by C.A.R. S\'a de Melo
(unpublished).
\end{references}
\end{document}